\documentclass[nofootinbib,preprint,aps,showpacs]{revtex4}

\usepackage{graphicx,epsfig}

\begin{document}

\title{Modification of the $\rho$ meson detected by 
low-mass electron-positron pairs in central Pb-Au collisions at 158$A$ GeV/$c$}

\def\rez{$^{(1)}$}
\def\gsi{$^{(2)}$}
\def\fra{$^{(3)}$}
\def\dub{$^{(4)}$}
\def\hei{$^{(5)}$}
\def\wei{$^{(6)}$}
\def\sun{$^{(7)}$}
\def\cer{$^{(8)}$}
\def\mpi{$^{(9)}$}
\def\bnl{$^{(10)}$}
\def\mun{$^{(11)}$}

\author{D.~Adamov\'a\rez, 
G.~Agakichiev\gsi, 
D.~Anto\'nczyk\gsi,
H.~Appelsh\"auser\fra, 
V.~Belaga\dub, 
J.~Bielcikova\hei,
P.~Braun-Munzinger\gsi,
O.~Busch\gsi, 
A.~Cherlin\wei, 
S.~Damjanovi\'c\hei, 
T.~Dietel\hei, 
L.~Dietrich\hei, 
A.~Drees\sun, 
W.~Dubitzky\hei,
S.\,I.~Esumi\hei, 
K.~Filimonov\hei, 
K.~Fomenko\dub,
Z.~Fraenkel\wei$^{\dagger}$, \footnotetext{$^{\dagger}$deceased} 
C.~Garabatos\gsi, 
P.~Gl\"assel\hei, 
J.~Holeczek\gsi, 
V.~Kushpil\rez, 
A.~Maas\gsi, 
A.~Mar\'{\i}n\gsi, 
J.~Milo\v{s}evi\'c\hei,
A.~Milov\wei, 
D.~Mi\'skowiec\gsi, 
Yu.~Panebrattsev\dub, 
O.~Petchenova\dub, 
V.~Petr\'a\v{c}ek\hei, 
A.~Pfeiffer\cer, 
J.~Rak\mpi, 
I.~Ravinovich\wei, 
P.~Rehak\bnl,
H.~Sako\gsi, 
W.~Schmitz\hei, 
S.~Sedykh\gsi, 
S.~Shimansky\dub, 
J.~Stachel\hei, 
M.~\v{S}umbera\rez, 
H.~Tilsner\hei, 
I.~Tserruya\wei, 
J.\,P.~Wessels\mun, 
T.~Wienold\hei, 
J.\,P.~Wurm\mpi, 
W.~Xie\wei, 
S.~Yurevich\hei, 
V.~Yurevich\dub \\
(CERES Collaboration)}

\affiliation{
\rez Nuclear Physics Institute ASCR, 25068 \v{R}e\v{z}, Czech Republic\\
\gsi Gesellschaft~f\"{u}r~Schwerionenforschung~(GSI),~D-64291~Darmstadt,~Germany\\
\fra Institut f\"{u}r Kernphysik der Universit\"{a}t Frankfurt,~D-60486 Frankfurt,~Germany\\
\dub Joint Institute for Nuclear Research, 141980 Dubna, Russia\\
\hei Physikalisches Institut der Universit\"{a}t Heidelberg, D-69120
Heidelberg, Germany\\
\wei Weizmann Institute, Rehovot 76100, Israel\\
\sun Department of Physics and Astronomy, State University of
New York--Stony Brook, Stony Brook, New York 11794-3800\\
\cer CERN, 1211 Geneva 23, Switzerland\\
\mpi Max-Planck-Institut f\"{u}r Kernphysik, D-69117 Heidelberg, Germany\\
\bnl Brookhaven National Laboratory, Upton, New York 11973-5000\\
\mun Institut f\"{u}r Kernphysik der Universit\"{a}t M\"unster, D-48149 M\"unster,~Germany\\
}

\begin{abstract}
  We present a measurement of $e^+e^-$ pair production in central Pb-Au collisions 
  at 158$A$ GeV/$c$.  
  As reported earlier, a significant excess of the $e^+e^-$ pair
  yield over the expectation from hadron decays is observed.  The improved
  mass resolution of the present data set, recorded with the upgraded CERES
  experiment at the CERN-SPS, allows for a comparison of the data with 
  different
  theoretical approaches. The data clearly favor a substantial in-medium
  broadening of the $\rho$~spectral function over a density-dependent shift of
  the $\rho$~pole mass.
  The in-medium broadening model implies
  that baryon induced interactions are the key mechanism to the observed
  modifications of the $\rho$ meson at SPS energy.

\end{abstract}

\maketitle 

\section{Introduction}

The masses of hadrons are created
dynamically by the strong interaction, when confinement forces quarks and
gluons to form color-neutral bound states. The generation of hadronic masses
is connected to spontaneous chiral symmetry breaking, a basic feature of
the vacuum structure of
Quantum-Chromo-Dynamics (QCD).  Evidently, the mechanism of chiral symmetry
breaking is of fundamental importance for the properties of matter in the
universe. However, the quantitative understanding of the dynamics in this
non-perturbative regime of QCD is still rather incomplete, and additional
information from experiment is essential. 

According to investigations of the non-perturbative properties of QCD on a
discrete space-time lattice a plasma of deconfined quarks and gluons (QGP)
should be 
formed at energy densities $\epsilon \geq 1$~GeV/fm$^3$.
Simultaneously with this deconfinement transition, chiral symmetry is expected
to be restored (see~\cite{lat} for a recent review).  In collisions of heavy
nuclei at high energies such energy densities are exceeded significantly and
there is by now strong, albeit indirect evidence for the formation of a QGP
(for recent reviews see~\cite{pbm,gyu,itz,sta}).  On the way to chiral symmetry
restoration in such matter, significant modifications of the properties of
hadrons are expected~\cite{pis,BR1}, such as of their mass and width or more 
generally of the hadronic spectral function.

The $\rho$~meson ($J^P=1^-$) is an ideal probe to investigate modifications of
such in-medium properties.  In a hot hadronic medium close to the phase
boundary, $\rho$~mesons are abundantly produced by annihilation of thermal
pions.  Due to its short lifetime ($c\tau = 1.3$ fm), the decay of the
$\rho~$meson occurs inside the medium, and spectral modifications may be
observable via the kinematic reconstruction of the decay products.  Finally,
its decay into lepton pairs provides essentially undisturbed information from
the hot and dense phase, because leptons are not subject to final state
rescattering in the strongly interacting medium.

Enhanced low-mass $e^+e^-$ pair production in nucleus-nucleus collisions at
full energy of the CERN-Super-Proton-Synchrotron (SPS) has been reported by
the CERES experiment.  In particular, in the mass region 0.2-0.6~GeV/$c^2$,
the measured di-lepton yield exceeds expectations from hadron decays by a
factor 2-3~\cite{ceres-sau,ceres-pbau,ceres-long}.  Even bigger enhancement
factors have been found at 40$A$~GeV/$c$~\cite{ceres-40}, albeit with
large statistical uncertainties.
At RHIC energies, enhancement factors similar to those observed at the top
SPS energy have been reported recently~\cite{phenix-electrons}.
 
Significant $\rho$~meson production via annihilation of thermal pions in the
hot and dense hadronic medium is a likely mechanism for enhanced electron pair
production. 
Implementing this mechanism,
substantial temperature and baryon density dependent
modifications of the $\rho$-spectral function~\cite{BR1,RW1,BR2} needed to be
considered to explain the mass spectrum of the pair enhancement.  However,
the detailed behaviour of the spectral function as chiral
symmetry is restored is still up to speculation. Quite different theoretical
approaches exist which could not be discriminated by the previous di-electron
data.

The NA60
Collaboration recently corroborated previous CERES findings and reported
a significant di-muon excess in nucleus-nucleus collisions
over the expectation from hadronic decays~\cite{na60-1}.
The NA60 measurement of
the di-muon excess in $^{115}$In-In collisions at 158$A$~GeV/$c$
favors models including significant broadening but no mass
shift of the $\rho$-spectral function~\cite{na60-1,rapp-hees}.

\section{Experiment and Data Analysis}
In this Letter, we present results on $e^+e^-$ pair production in central
$^{208}$Pb-$^{197}$Au 
collisions at 158$A$~GeV/$c$. The data have been recorded by the CERES
experiment at the SPS in the year 2000~\cite{ana-qm}.  
Typically, $10^6$ lead ions per 5.2~s
extraction cycle were focused on 13 thin gold targets aligned along the beam
line (25~$\mu$m each, totalling 1.2\% of a nuclear interaction length).  The
interaction vertex was reconstructed using charged particle track segments
from two silicon drift detectors (SDD) placed 10.4 and 14.3 cm downstream of
the target.  Electrons are identified by their ring signature in two RICH
detectors, which are blind to hadrons below $p\approx 4.5$ GeV/$c$ 
($\gamma_{\rm thresh} \approx 32$).  The
experimental setup was upgraded by a downstream radial drift Time Projection
Chamber (TPC) which is operated inside an inhomogeneous magnetic field with a
radial component of up to 0.75~T.  
Employing tracking information from the
TPC, the mass resolution of the spectrometer was improved to $\Delta m/m = 3.8
\%$ in the region of the $\phi$~meson mass~\cite{ana-qm}.  
The resolution has been determined by a Monte Carlo procedure 
where simulated tracks were embedded into real data events. 
This method 
provides a detailed simulation of the TPC response to charged particles 
in a realistic track density environment. 
The same Monte Carlo describes very well the experimentally observed peak width 
of the $K^0_s$ reconstructed in the $ \pi^+\pi^-$ channel 
($\sigma_{K^0_s\rightarrow \pi^+\pi^-} \approx 15$~MeV/$c^2$). 
The TPC also
provides additional electron identification via measurement of the specific
energy loss ${\rm d}E/{\rm d}x$ with a resolution of about 10\%.  The
spectrometer provides full azimuthal acceptance in the pseudorapidity range
$2.1 < \eta < 2.65$.

The present results are based on an analysis~\cite{sergey,alex,oli} of 
25 million Pb-Au

events, selected at a centrality of $\sigma/\sigma_{\rm
  geo}=7~\%$~\cite{darek-qm}.  SDD track segments are matched to charged
particle tracks in the TPC, where the deflection in the magnetic field
determines the momentum with a resolution of $\Delta p/p \approx \left((2\%)^2
  +(1\%\cdot p ({\rm GeV/}c) )^2\right)^{1/2}$~\cite{ana-qm}.  Combined
electron information from cuts on the ring quality in the RICH detectors and
TPC d$E$/d$x$ leads to a pion suppression of typically $4~\cdot10^4$ at 67\%
electron efficiency~\cite{sergey}.

A set of cuts is applied to the track candidates
to minimize the amount of combinatorial background from unrecognized
Dalitz pairs and conversions, making use of their characteristic
decay topology.
Electron and positron tracks are rejected if
their d$E$/d$x$ significantly exceeds that of a single track in both SDDs,
indicating an unresolved close pair. Also, electron and positron tracks are
rejected if a soft track with opposite charge and electron-like d$E$/d$x$
is found in the TPC at small angular separation. 

The remaining electron and positron tracks are then combined into pairs.
To further suppress combinatorial background, tracks which form
pairs with opening angle smaller than 35 mrad are treated as recognized
conversions or Dalitz pairs and are not used for further pairing. 
Finally, a low transverse momentum cut
of 0.2 GeV/$c$ is applied to all electron and positron tracks.

The single-electron 
reconstruction efficiency has been determined by a Monte-Carlo (MC) procedure
where simulated tracks are embedded into real raw data events. The subsequent
analysis of the MC sample includes all cuts and methods as applied to the real
data.  
The final pair reconstruction efficiency $\epsilon_{\rm ee}$ is typically 
14\%. 
It depends on the polar angles of the single tracks and
(slightly) on centrality. 
The efficiency correction is performed by assigning a weight $w_{\rm ee}$ to each pair
which is the inverse of the pair reconstruction efficiency $\epsilon_{\rm ee}$.
The pair reconstruction efficiency is calculated from the single track efficiencies,
using the parametrized dependencies on track polar angle $\theta$ and charged 
particle multiplicity $N_{\rm ch}$:
\begin{equation}
w_{\rm ee} =
\frac{1}{\epsilon_{\rm ee}}=
\frac{1}{\epsilon_{\rm track1}(\theta_1, N_{\rm ch})\cdot
         \epsilon_{\rm track2}(\theta_2, N_{\rm ch})}.
\end{equation}

Based on a systematic variation of the cut values in the MC, the systematic 
uncertainty on the pair reconstruction efficiency is estimated to 8.4\%. 
\begin{figure}[h]
\begin{center}
\includegraphics[width=7.5cm]{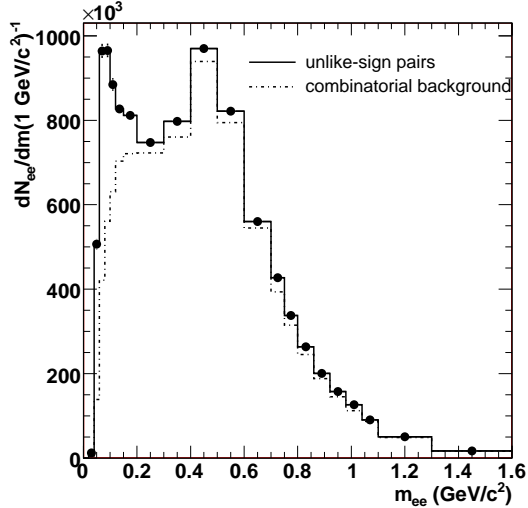}
   \caption{Unlike-sign pair yield (histogram) and
     combinatorial background (dashed curve). See text for explanation.}
\label{fig1}
\end{center}
\end{figure}

The remaining combinatorial background is estimated to equal the number of
like-sign lepton pairs from the same event. 
The limited statistics of the like-sign pair sample adds a significant 
statistical error to the final result. 
Alternatively, a mixed-event technique can be applied, where
unlike-sign pairs are formed from electron and positron tracks of different
events and the corresponding mixed-event spectrum is normalized to
the same-event like-sign yield. 
This approach reduces considerably the statistical bin-to-bin fluctuations
but bears a 
statistical and 
systematic uncertainty due to the normalization procedure.
The normalization constant is determined by the ratio of the 
mixed-event spectrum to the same-event like-sign spectrum. The statistical
accuracy of the normalization is limited by the number of counts in the 
same-event like-sign spectrum and yields about $4\cdot 10^{-3}$. 
The uncertainty in the normalization of the 
background 
has been included in
the systematic error of the final dilepton yield.
The magnitude of this contribution after background subtraction depends 
on the signal-to-background ratio and yields typically 8.8\% at $S/B$ =~1/22 
(see below).

We found that the final
results 
using the like-sign and the mixed-event sample are in good agreement within statistical
errors~\cite{sergey}.  Small deviations are only visible at invariant masses
below 0.1 GeV/$c^2$.  This is caused by limitations of the two-ring
separation, which are only present in the same-event sample.  To account for
this effect, which is also present in the 'true' unlike-sign combinatorial
background, we have used the same-event like-sign background estimate for
masses below 0.2~GeV/$c^2$.  No statistical limitation is imposed by this
procedure, since the signal-to-background ratio is very good in this mass
region.  For masses greater than 0.2~GeV/$c^2$, the normalized mixed-event
unlike-sign sample is used for background subtraction. 
The resulting background distribution 
and the unlike-sign signal pair spectrum after efficiency correction
are shown in Fig.~\ref{fig1}.

Below $m_{\rm ee}$~=~0.2~GeV/$c^2$ the $\pi^0$-Dalitz contribution
is clearly visible. 
The raw net yield 
in this mass range after subtraction of the combinatorial background contains
$6114\pm 176$ electron-positron 
pairs at a signal-to-background ratio 
$S/B$~=~1/2.
At masses larger than 0.2~GeV/$c^2$, the number of 
pairs contributing to the signal is 
$3115\pm 376$
at
$S/B$~=~1/22.

After background subtraction, the $e^+e^-$ pair yield is normalized to the
total number of events and to the average charged particle multiplicity
$\langle N_{\rm ch}\rangle$ in the spectrometer acceptance. 
The average charged particle multiplicity has been
determined by the number of tracks in the SDD. 
For the 7\% most central events
we obtain 
$\langle N_{\rm ch}\rangle = 177 \pm 14$ (syst.)
in $2.1 < \eta < 2.65$. 
The systematic error on $\langle N_{\rm ch}\rangle$ adds
a contribution of 8\% to the total systematic error on the pair yield
per charged particle.

The total systematic error of the data is given 
by (i) the uncertainty of the efficiency determination (8.4\%), (ii) 
the normalization of the 
mixed-event background ($0.004 \cdot B/S$, on average 8.8\%) and (iii) 
the determination of the charged particle 
multiplicity (8\%). 
These contributions
add up to an average of 14.6\%, however, note that contribution (ii) differs
bin-by-bin, depending on the local signal-to-background ratio, and applies only for
$m_{\rm ee} > 0.2$~GeV/$c^2$. 

\section{Results and Discussion}

\begin{figure}
\begin{center}
   \includegraphics[width=7.5cm]{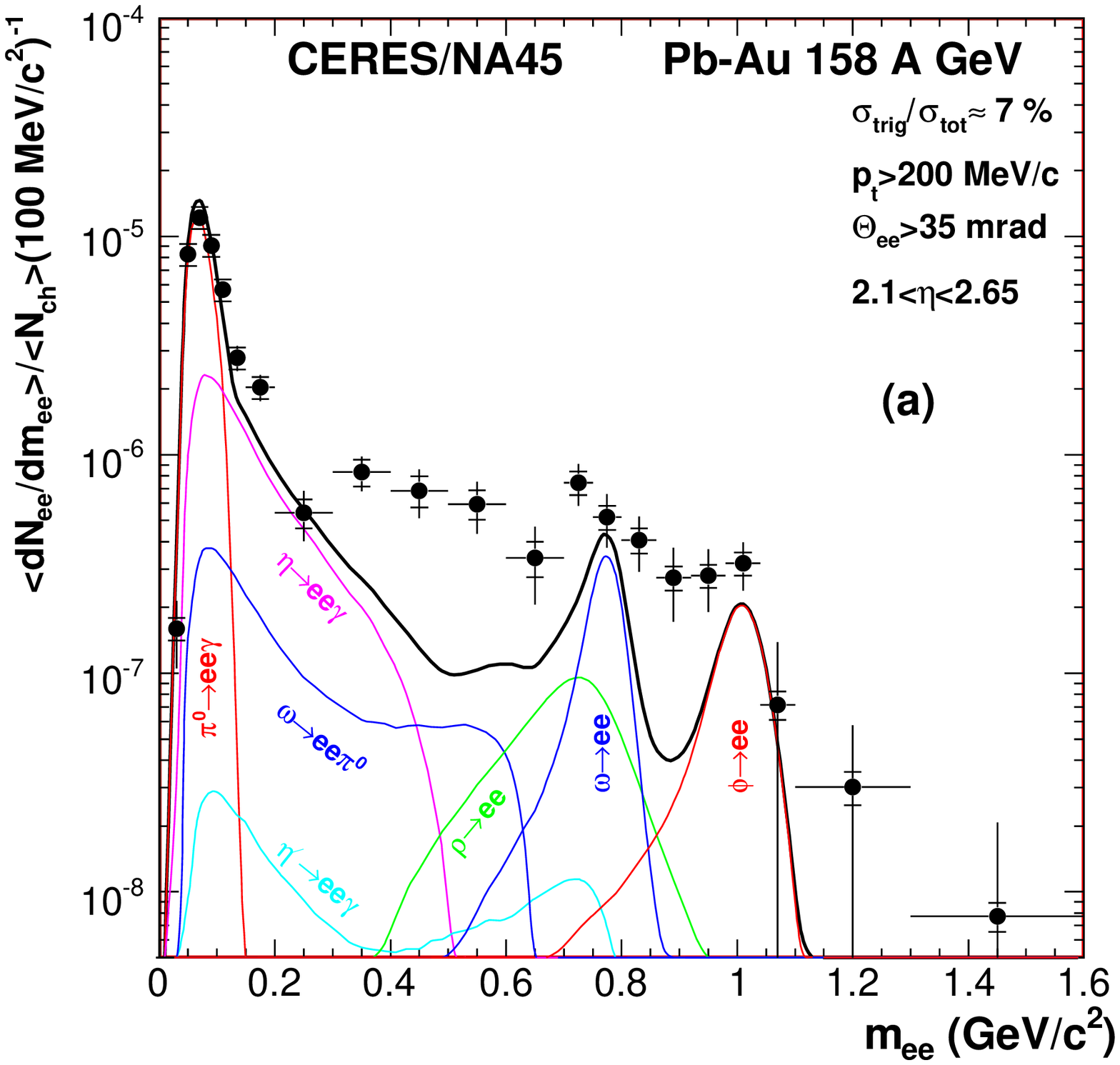}   
   \includegraphics[width=7.5cm]{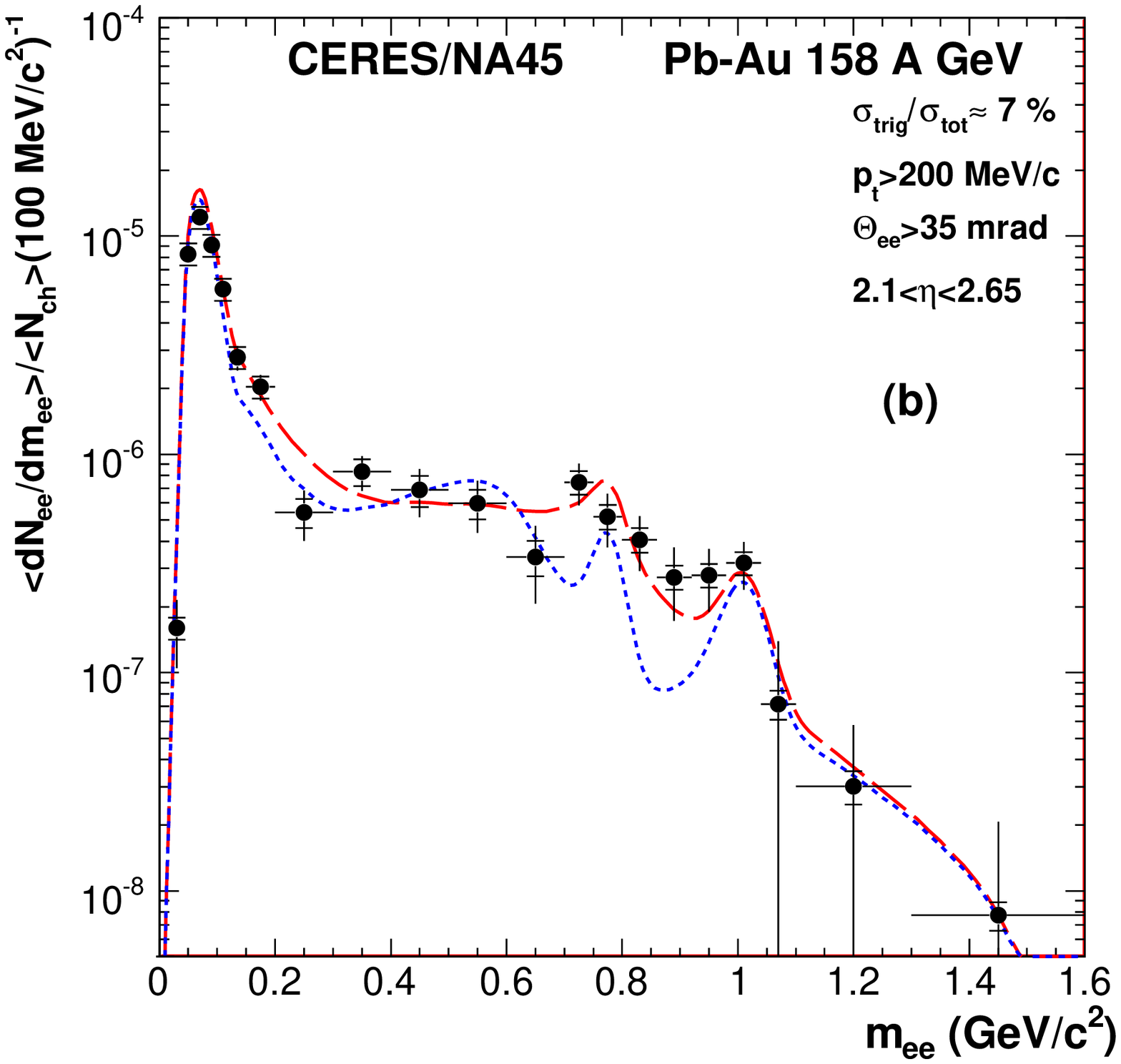}
   \caption{a: Invariant $e^+e^-$ mass spectrum compared
to the expectation from hadronic decays. b: The same data
compared to calculations including a dropping $\rho$ mass (dashed)
and a broadened $\rho$-spectral function (long-dashed). 
Systematic
errors are indicated by horizontal ticks.}
\label{fig2}
\end{center}
\end{figure}

The $e^+e^-$ invariant mass distribution after 
efficiency correction, combinatorial
background subtraction and normalization
is shown in Fig.~\ref{fig2}~(a). 
Also shown 
is the 'hadronic cocktail' which comprises the yield from hadronic decays in
A-A collisions after chemical freeze-out 
(see~\cite{ceres-long})\footnote{for the $\phi$~meson we 
use 70\% of the thermal model yield,
in accordance with measurements~\cite{ceres-phi}. 
The calculation of the present cocktail includes also the
recently improved branching ratio for 
$\omega \rightarrow \pi^0 e^+ e^-$~\cite{cockt-update}.}.
In the mass range $0.2 <m_{\rm ee} < 1.1$~GeV/$c^2$, the data are enhanced 
over the 
cocktail by a factor
$2.45 \pm 0.21~{\rm (stat)} \pm 0.35 ~{\rm (syst)} \pm 0.58~{\rm (decays)}$.
The last error arises from the systematic uncertainty in the cocktail
calculation. The enhancement is most pronounced in the mass region $0.2 <
m_{\rm ee} < 0.6$~GeV/$c^2$, in agreement with earlier findings.  In contrast
to previous CERES results, the improved mass resolution of the upgraded
spectrometer provides access to the resonance structure in the $\rho / \omega$
and $\phi$ region. A quantitative study of $\phi$~meson production in the
$e^+e^-$ and $K^+K^-$ channels can be found in~\cite{ceres-phi}.

In Fig.~\ref{fig2}~(b) the data are compared with a model
approach implying enhanced di-lepton production via thermal pion
annihilation and a realistic space-time evolution~\cite{rapp-calc}.
The calculated di-lepton yield was filtered by the CERES acceptance
and folded with the experimental resolution.
Temperature and baryon-density dependent modifications of the $\rho$-spectral function have been taken into account: 
the dropping mass scenario which assumes a
shift of the in-medium $\rho$ mass~\cite{BR1,BR2}, and the broadening scenario
where the $\rho$-spectral function is smeared due to coupling to the hadronic
medium~\cite{RW1,rapp-hees}. 
The calculations include as well contributions from QGP,
the Drell-Yan process, and 4-pion annihilation with chiral mixing. 
The calculations for both spectral functions describe the enhancement
reasonably well for masses below 0.7~GeV/$c^2$. 
In the resonance region, however, there is a notable difference between
the calculations.
In particular in the mass region
between the $\omega$ and the $\phi$, the data clearly favor the broadening
scenario 
over the dropping mass scenario.
\begin{figure}
\begin{center}
   \includegraphics[width=7.cm]{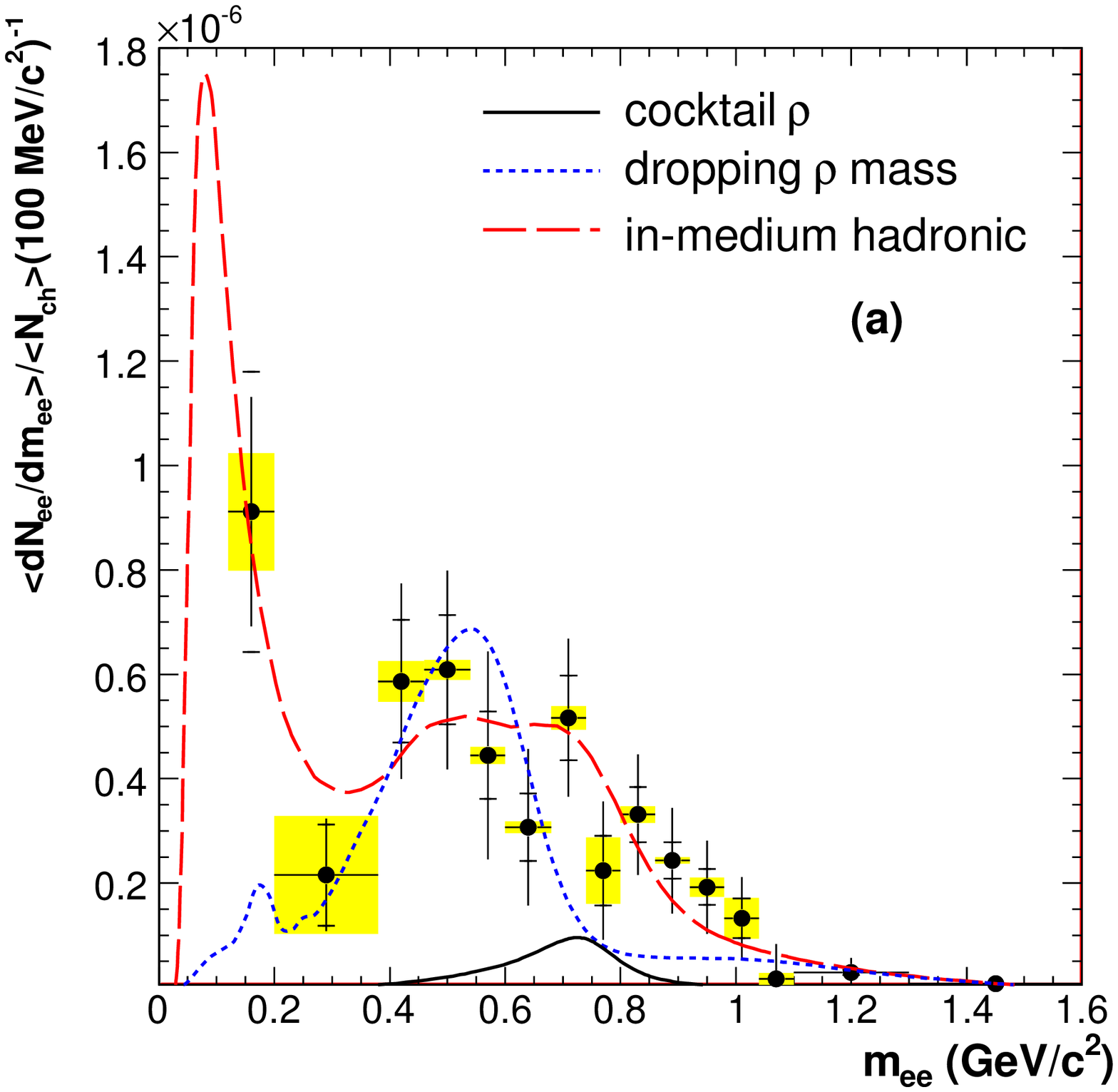}
   \includegraphics[width=7.cm]{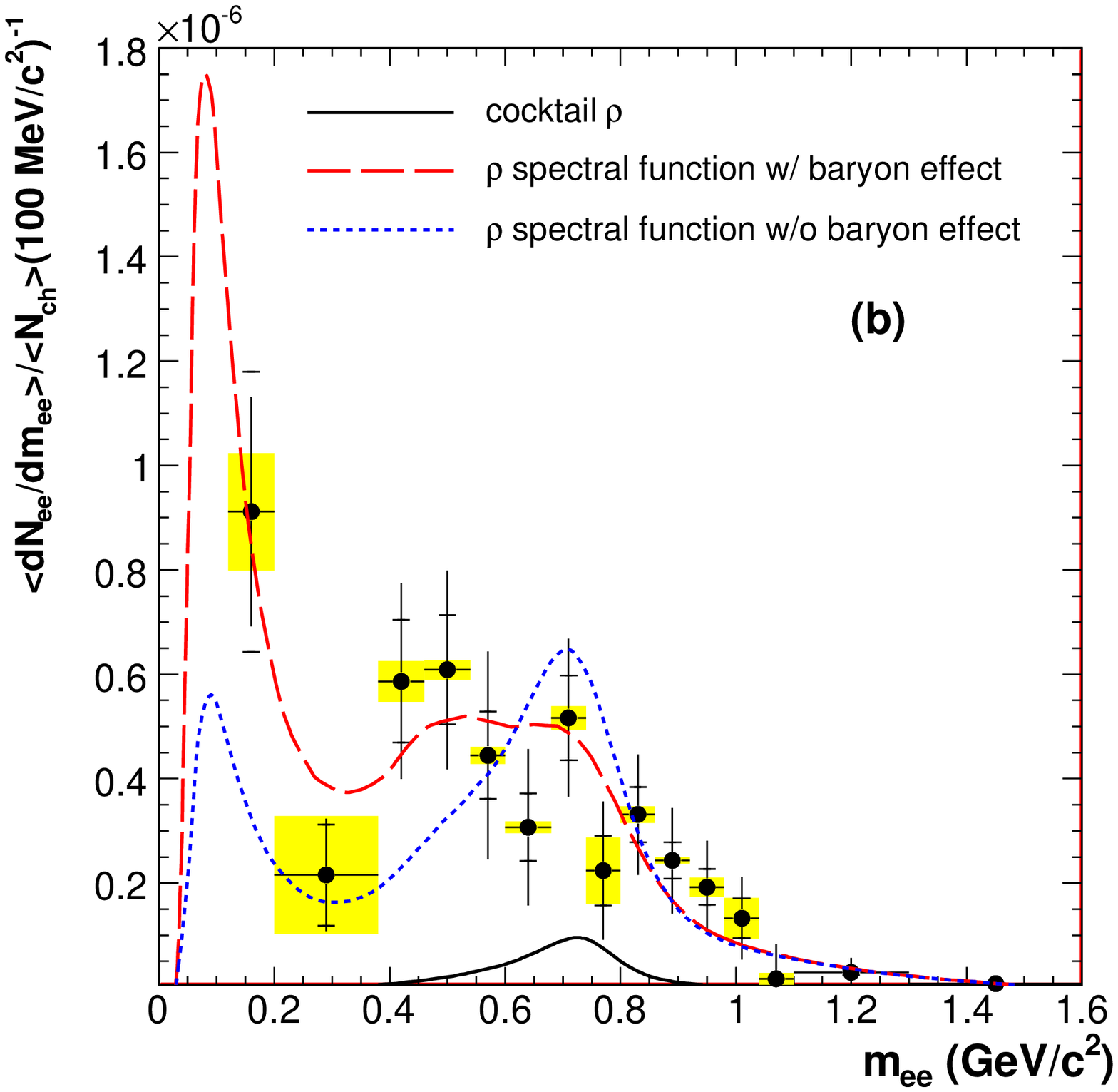}
   \caption{$e^+e^-$ pair yield after subtraction of the hadronic
     cocktail. In addition to the statistical error bars,
   systematic errors of the data 
	(horizontal ticks) and the systematic uncertainty of the subtracted 
	cocktail (shaded boxes) are indicated.
     The broadening scenario (long-dashed line) is compared to a calculation
     assuming a density dependent dropping $\rho$ mass (dotted line in (a)) 
     and to a broadening scenario excluding baryon effects (dotted line
     in (b).}
\label{fig3}
\end{center}
\end{figure}

In order to exhibit the shape of the in-medium contribution,
we subtract the hadronic cocktail (excluding the
$\rho$ meson) from 
the data (Fig.~\ref{fig3}). 
The vacuum $\rho$-decay contribution to the data 
(``cocktail $\rho$'', solid line in Fig.~\ref{fig3}) 
is completely negligible compared to the measurements.
The excess data exhibit a very broad structure reaching very low masses
and exceed the vacuum $\rho$ contribution by a factor $10.6\pm1.3$. 
The data are compared to model calculations of the in-medium di-electron
production. These are normalized, like the measured yield, to the number
of charged particles.
Note that the model calculations give absolute pair yields (in terms of
charged particle numbers) and there is no freedom of adjustment.
Yield and spectral shape
are well described by the broadening scenario but are not
consistent{\footnote{Recently, Brown et al.~\cite{newbrownrho} have 
advocated a
  different view in which their scaling is not directly related anymore to the
  shape of the low mass di-lepton spectrum.}} with a dropping $\rho$ mass
(Fig.~\ref{fig3}~(a)). 
While the dropping mass calculation yields a rather narrow distribution
with a peak around 0.5~GeV/$c^2$ the measured excess is spread over a 
significantly wider mass range. 
Below 0.2~GeV/$c^2$, the large errors arising from the subtraction of large
numbers in the $\pi^0$ Dalitz region do not allow for a definite conclusion.
However, the trend indicates a further increase of the in-medium contribution
towards the photon point.

A $\chi^2$-analysis of the data 
in the mass region 
$0.12 < m_{\rm ee} < 1.1$~GeV/$c^2$ ($\rm{dof} = 13$)
with respect to the model calculations in Fig.~\ref{fig3} results in 
$\chi^2_{\rm IMH}=10.6$ ($P_{\rm IMH (stat) } =64.4\%$)
for the in-medium hadronic spectral function approach and $\chi^2_{\rm DRM}=33.1$ 
($P_{\rm DRM (stat)} =0.0017\%$) for the dropping $\rho$ mass scenario, 
if only statistical errors are considered.
To judge how well the data resemble the calculations shown in 
Fig.~\ref{fig3}~(a)
including systematic uncertainties in data and cocktail, a Monte Carlo procedure
has been employed. 
In this procedure, the model curves have been used as input to 
generate simulated spectra, assuming statistical and systematic uncertainties
as in the real data.
The total systematic error in each mass bin has been calculated 
by adding in quadrature 
the systematic error contributions to data and cocktail.
The resulting total systematic error is interpreted as a Gaussian 
standard deviation of a
coherent up- or downward shift of all data points in the spectrum.
Statistical point-to-point fluctuations have been added
according to the statistical error bars of the real data.
For each of the generated spectra,
the $\chi^2$-value 
with respect to the input model
curve has been calculated.
Finally,
the probability to obtain a $\chi^2$-value which is 
larger than the one observed in the data has been determined for each of the two
model curves. We obtain 
$P_{\rm IMH} = 81.1\%$ for the in-medium hadronic spectral function approach
and $P_{\rm DRM} = 10.1\%$ for the dropping $\rho$ mass scenario, the latter 
implying that the dropping $\rho$ mass scenario can be excluded on the
$1.6~\sigma$-level only. We note that, despite the statistical limitations of
the present data, the discrimination power among the present model calculations
is predominantly limited by systematic uncertainties.

A more detailed view may be derived 
by comparing the gross features of the data 
in Fig.~\ref{fig3}~(a) to the model calculations, see Table~\ref{tab1}. 
The systematic errors on these quantities have been estimated by shifting each
data point up or down by one standard deviation of its total systematic error.
The mean
values of the mass distributions of both calculations are in very good agreement with
the measured excess data. The integrated yield in $0.12<m_{\rm ee}<1.1$~GeV/$c^2$ 
agrees well with the in-medium hadronic spectral function approach, however,
the systematic uncertainty does not exclude the dropping $\rho$ mass scenario
either. In contrast, a comparison of the RMS widths clearly favors the
in-medium hadronic spectral function approach, 
as the width of the data distribution is quite insensitive to systematic errors 
in scale of both data and cocktail.
On this note, it is the large width observed in the data which drives the 
discrimination power among the model calculations.

\begin{center}
\begin{table}
\caption{Di-electron excess in $0.12<m_{\rm ee}<1.1$ GeV/$c^2$ 
compared to model calculations.}
~
\begin{tabular}{c|c|c|c}
\hline \hline
~ & Data 			& ~In-medium~ & ~Dropping~ \\
~ & ~($\pm$ stat.) ($\pm$ syst.)~ & ~hadronic~  & ~$\rho$ mass~ \\ 
\hline
Mean (GeV/$c^2$)  &~0.54$\pm$0.07$\pm$0.01~ & 0.54 & 0.55 \\
Yield ($10^{-6})$ & 3.58$\pm$0.42$\pm$1.01 & 3.88 & 2.41 \\
RMS (GeV/$c^2$)   & 0.26$\pm$0.02$\pm$0.01 & 0.25 & 0.18 \\
\hline \hline
\end{tabular}
\label{tab1}

\end{table}
\end{center}

The agreement of the data with the broadening scenario 
is strong evidence that the
resonance structure of the $\rho$ meson 
is significantly modified
in the hot
and dense medium~\cite{rapp-hees,qh-dual}.
That $\rho$-related pair production is indeed a manifestation of the 
hot and dense matter created is supported by observing a particular mechanism 
at work which plays a dominant role in the hadronic spectral function
approach:
The strong coupling to baryons which adds
strength to the di-electron yield at low masses~\cite{RW1}. The importance of
this mechanism is demonstrated in Fig.~\ref{fig3}~(b), where the data
are compared to in-medium hadronic spectral function calculations 
with and without baryon-induced
interactions~\cite{rapp-hees,rapp-calc}. 
The calculations differ most in the mass range
below 0.5~GeV/$c^2$, which is accessible with good efficiency by the present
$e^+e^-$ data. The calculation omitting baryon effects 
falls short of
the data for masses below 0.5~GeV/$c^2$ while inclusion of baryon interactions
describes the measured low-mass yield very well. 
This is strong evidence that the observed modifications of the $\rho$-spectral
function are foremost due to interactions with the dense baryonic medium.

It has been demonstrated that baryon-driven medium modifications lead to
a low-mass di-electron spectrum which is very similar to 
the di-electron rate from lowest order perturbative $q\overline{q}$ 
annihilation~\cite{qh-dual}. 
Inspired by this apparent emergence of quark-hadron duality 
at low masses, Gallmeister et al. performed a
calculation assuming that di-electron
production via $q\overline{q}$
annihilation at $T_c$ is dominant~\cite{kaempfer}. 
That this phenomenological approach describes the shape of the measured
distribution (Fig.~\ref{fig3}) so well may be taken as an indication 
of chiral symmetry restoration which is implied in quark-hadron duality.

In cold nuclear matter, medium modifications not only of 
$\rho$ but also of $\omega$ and $\phi$ mesons
have been observed~\cite{trnka,naruki,muto} (see~\cite{metag} for a recent 
review). 
Such modifications may also be
present in the medium created in high-energy heavy-ion collisions. 
It should be noted however that the dominance 
of thermal two-pion annihilation to the total di-lepton yield 
via the $\rho$-channel disfavors a significant 
contribution from $\omega$ and $\phi$ to the observed di-lepton
excess and its spectral distribution.

In conclusion, the present $e^+e^-$ data with improved mass resolution in the
resonance region 
favor present models including a strong broadening of 
the $\rho$-spectral 
function in a hot and dense hadronic medium over a density dependent
$\rho$ mass shift.  Moreover, the 
$e^+e^-$ data at low pair
mass and transverse momentum allow to test the relevance of baryonic effects
to the modification of the $\rho$-spectral function.  
In comparison with models, the present CERES data suggest that baryonic interactions
are important to explain the observed di-lepton 
yield at low masses.

\section{Acknowledgements} \nonumber

This work was supported by GSI, the German BMBF, the Virtual Institute VI-SIM
of the German Helmholtz Association, the Israel Science Foundation,
the Minerva Foundation and by the Grant Agency and Ministry of Education
of the Czech Republic. 
We wish to thank Ralf Rapp for calculations and 
numerous discussions.


\begin{thebibliography}{9}
\bibitem{rapp-calc} all calculations were performed by R. Rapp (private
  communication) based on~\cite{qh-dual}.
\end{thebibliography}
\end{document}